\documentclass[]{spie}  

\usepackage{amsmath,amsfonts,amssymb}
\usepackage{graphicx}
\usepackage[colorlinks=true, allcolors=blue]{hyperref}

\title{The MICADO first light imager for the ELT:
the PSF Reconstruction Software}

\author[a]{Andrea Grazian}
\author[b]{Elisa Portaluri}
\author[a]{Matteo Simioni}
\author[a]{Carmelo Arcidiacono}
\author[a]{Marco Gullieuszik}
\author[c,d]{Johanna Hartke}
\author[e]{Daniel Jodlbauer}
\author[g]{Fernando Pedichini}
\author[g]{Roberto Piazzesi}
\author[g]{Piero Vaccari}
\author[a]{Benedetta Vulcani}
\author[f]{Roland Wagner}
\author[a]{Anita Zanella}
\affil[a]{INAF-Osservatorio Astronomico di Padova,
  Vicolo dell’Osservatorio 5, I-35122, Padova, Italy}
\affil[b]{INAF-Osservatorio Astronomico d’Abruzzo,
  Via Mentore Maggini, s.n.c. I-64100, Teramo}
\affil[c]{Finnish Centre for Astronomy with ESO (FINCA),
  University of Turku, Finland}
\affil[d]{Tuorla Observatory, Department of Physics and Astronomy, FI-20014,
  University of Turku, Finland}
\affil[e]{Industrial Mathematics Institute, Johannes Kepler University Linz,
  Altenberger Strasse 69, 4040 Linz, Austria}
\affil[f]{RICAM - Johann Radon Institute for Computational and
  Applied Mathematics, Altenberger Strasse 69, 4040 Linz, Austria}
\affil[g]{INAF - Osservatorio Astronomico di Roma, Via Frascati 33,
  I-00078, Monte Porzio Catone, Italy}

\authorinfo{Further author information: (Send correspondence to A. Grazian)\\
A. Grazian: E-mail: andrea.grazian@inaf.it, Telephone: 0039-049-8293465}

\begin{document} 
\maketitle

\begin{abstract}
MICADO is the first-light camera of the ESO ELT, allowing NIR imaging
and long-slit spectroscopy assisted by adaptive optics. MICADO is now
entering its construction phase, and the software for data reduction
is reaching an adequate maturity level. The PSF Reconstruction (PSF-R)
of MICADO is a software tool for the blind derivation of the PSF, only
using adaptive optics telemetry data. An update of the status of the
PSF-R service is provided here. The PSF-R prototype has been tested on
ERIS@VLT data in order to check the reconstruction of on- and off-axis
PSFs. The on-axis PSF-R is accurate at a few percent level on Strehl,
FWHM, Encircled Energy, and half light radius, while for the off-axis
case the match is within 10-15\% at a distance of half isoplanatic
angle. The first version of the workflow for the PSF-R pipeline has
been developed and verified using the latest release of the ESO data
processing system. A set of simulations has been implemented on
the morphological analysis of distant galaxies, showing that the
accuracy of the PSF-R matches the goals needed to study their
morphology. In summary, the PSF-R team is on the right track towards
the ELT first light.
\end{abstract}


\section{INTRODUCTION}
\label{sec:intro}

State-of-the-art Adaptive Optics (AO) observations with big telescopes
is an old dream \cite{ragazzoni06} that is close to reality in the
present days. The Multi-Adaptive Optics Imaging Camera for Deep
Observations (MICADO) is the first-light camera of the European
Southern Observatory (ESO) Extremely Large Telescope (ELT). This
workhorse facility will allow, within a few years from now, to obtain
imaging and long-slit spectroscopy assisted by AO at near-infrared
wavelengths \cite{davies16}.

In the course of fulfilling the final design review, MICADO is now
successfully entering its assembly and construction phase. At the same
time, the software for the data reduction of MICADO is going to reach
a progressively adequate maturity level. Within the data reduction
pipeline of MICADO, the PSF Reconstruction (PSF-R \cite{grazian22}) is
a software tool for the blind derivation, in post processing, of the
PSF of a given observation, starting only from AO telemetry, without
accessing the focal plane data \cite{wagner18}.

It will be the first time, within all ESO telescopes, that such a
service is offered to the scientific community. The PSF-R software
tool will work both for Single Conjugate (SCAO) and Multi Conjugate
Adaptive Optics (MCAO), the latter allowed by the MORFEO
\cite{ciliegi22} module that is foreseen to be available approximately
2 years after the first light of MICADO. An update on the status of
the PSF-R service tool of MICADO is provided here.

\section{METHODS}
\label{sec:method}

\begin{figure}[ht]
\begin{center}
\begin{tabular}{c}
\includegraphics[width=16cm]{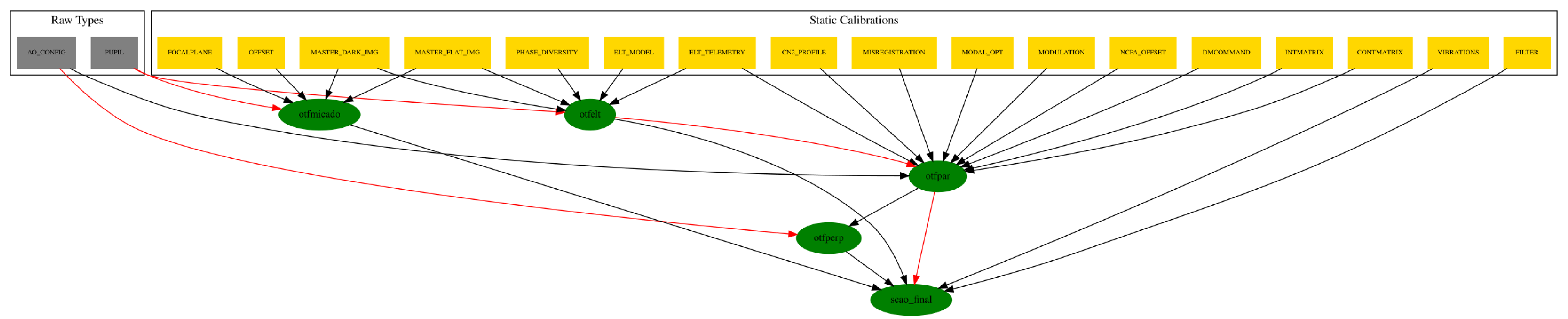}
\end{tabular}
\end{center}
\caption[EDPS Workflow]
{\label{fig:edps}
The EDPS workflow of the PSF-R software for MICADO, split in two panels.
Grey boxes indicate the raw input data, yellow boxes
indicate the calibration files, while green ellipses show the
five recipes. Black and red arrows indicate data processing.}
\end{figure}

The PSF-R software is organized in five main recipes that have been
further divided into atomic functions with the aim of data
classification, organization, handling, and processing. The recipes
are orchestrated by a high level software, the ESO Data Processing
System (EDPS \cite{Freudling24}). The EDPS workflow of the PSF-R
software for MICADO is show in Fig. \ref{fig:edps}. One recipe and
four functions have been already developed in C code and run through
EDPS, and their debugging is currently ongoing.

The scientific evaluation of the PSF-R software is also currently ongoing,
with particular emphasis on the physical properties (e.g. size,
stellar mass, Star Formation Rate) of star forming regions (clumps) of
high-redshift galaxies (Simioni et al., SPIE, this conference).

\section{RESULTS}
\label{sec:results}

While waiting for the first slopes and matrices from MICADO, in
preparation for the Preliminary Acceptance in Europe (PAE) phase, the
present version of the PSF-R prototype software has been already
tested on real data from working SCAO systems, e.g. LBT/SOUL
\cite{pinna16} and VLT/ERIS \cite{davies23}.

\subsection{PSF Reconstruction with LBT/SOUL data}

\begin{figure}[ht]
\begin{center}
\begin{tabular}{c}
\includegraphics[width=16cm]{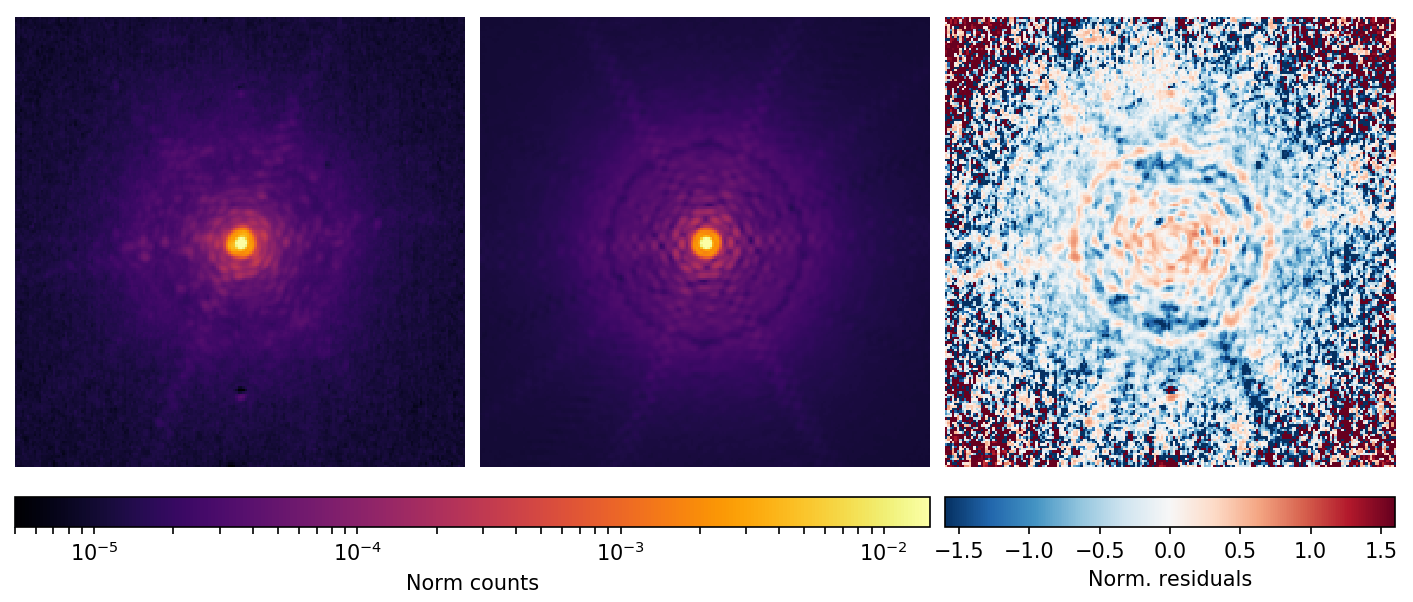}
\end{tabular}
\end{center}
\caption[LBT/SOUL PSF-R]
{\label{fig:soul}
SOUL+LUCI PSFs. Left panel: Observed PSF. Central panel:
Reconstructed PSF. Right panel: Normalized residuals between observed
and reconstructed PSF. The field of view (FoV) is 8 arcsec wide
for all the three panels.}
\end{figure}

The PSF-R software tool of MICADO has been successfully applied to
LBT/SOUL+LUCI data (Fig. \ref{fig:soul}) in the case of
a bright on-axis star observed in SCAO mode. With this configuration, an
accuracy of 2-4\% has been reached on Strehl Ratio (SR), Full Width at
Half Maximum (FWHM), and Encircled Energy (EE) \cite{Simioni20,Simioni22}.
A technology readiness level (TRL) of 7 has then been reached \cite{grazian22}.

\subsection{PSF Reconstruction with VLT/ERIS data}

ERIS@VLT data have been acquired purportedly to check for the
reconstruction of on- and off-axis PSFs. Encouraging results have been
obtained by applying the PSF-R software to the ERIS@VLT data of an
asterism of three bright stars with separation less than 16 arcsec.

\begin{figure}[ht]
\begin{center}
\begin{tabular}{c}
\includegraphics[width=16cm]{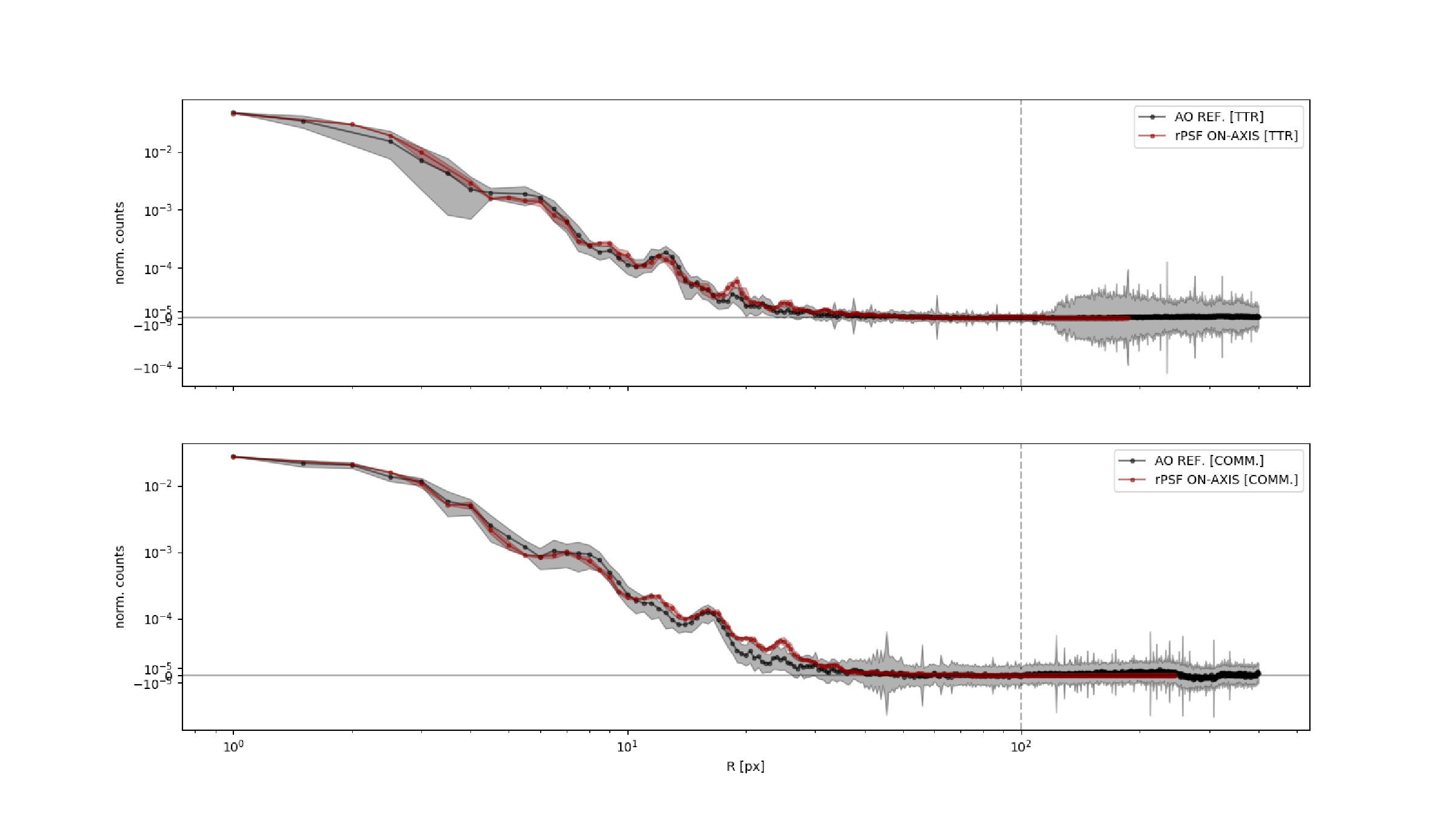}
\end{tabular}
\end{center}
\caption[ERIS PSF-R]
{\label{fig:eris}
ERIS@VLT observations of two bright stars (on-axis). The black lines
show the radial profiles of the observed stars, while the red curves
show the reconstructed PSFs. Grey bands indicate the uncertainties (1
$\sigma$) of the observed profiles.}
\end{figure}

The on-axis PSF reconstruction (Fig. \ref{fig:eris}) is accurate at
$\sim 2\%$ level, while for the off-axis case the match is within
approx 10-15\% at a distance of half isoplanatic angle, adopting as
comparison the standard metrics (SR, FWHM, EE, and half light radius
of the PSF). Additional work is currently in progress (using the
method described in Ref.~\citenum{wagner23}). See also the contribution by
Matteo Simioni in this Proceeding.

\section{CONCLUSIONS}

The first version of the workflow for the PSF-R pipeline has been
developed and verified using the latest release of the ESO data
processing system (EDPS). In parallel, the development of individual
recipes of the PSF-R pipeline has been started. Testing of the PSF-R
software tool is currently ongoing using ERIS@VLT imaging of bright
star asterisms as a benchmark. The SR, FWHM, EE of the on-axis star
has been reproduced at 2\% accuracy by the reconstructed PSF tool. The
method has been successfully applied also to an off-axis star with a
reduction of the performances on the recovery of SR, FWHM, EE of just
$\sim$10-15\%. Further progresses are expected in the future by the
PSF-R Team of MICADO for the reliability of the off-axis PSF
reconstruction.

In order to check the reliability of the reconstructed PSFs, a set of
simulations has been implemented on the morphological analysis of
distant galaxies. The scientific evaluation of the PSF-R products
shows that the accuracy of the reconstructed PSF, reached with the
preliminary version of the PSF-R software of MICADO, matches the goals
needed to study the morphological parameters of extra-galactic
objects. In summary, the PSF-R team of MICADO is on the right track
towards the first light of ELT and its initial scientific exploitation.

\acknowledgments

This work has been partly supported by INAF through the Math,
ASTronomy and Research (MAST\&R), a working group for mathematical
methods for high-resolution imaging.

MS and AG acknowledge support from Bando Ricerca Fondamentale INAF 2023
(Ob. Fu. 1.05.23.04.05, P.I. Simioni).

\bibliography{report} 
\bibliographystyle{spiebib} 

\end{document}